\input{aipcheck}

\documentclass[
    ,final            
  ]
  {aipproc}

\layoutstyle{6x9}

\newcommand {\cK}{{\cal K}}

\newcommand {\cM}{{\cal M}}
\newcommand {\cN}{{\cal N}}

\newcommand {\cS}{{\cal S}}

\def\a{\alpha}

\def\z{\zeta}

\def\F{\Phi}
\def\J{\Psi}
\def\L{\Lambda}

\def\S{\Sigma}
\def\U{\Upsilon}

\newcommand{\pa}{\partial}                           

\newcommand{\be}{\begin{equation}}
\newcommand{\ee}{\end{equation}}
\newcommand{\bea}{\begin{eqnarray}}
\newcommand{\eea}{\end{eqnarray}}
\newcommand{\non}{\nonumber}
\begin{document}

\title{Projective superspace and 
hyperk\"ahler sigma models on cotangent bundles of Hermitian
symmetric spaces}

\classification{11.10.Lm, 11.30.Pb}
\keywords      {hyperk\"ahler manifold, sigma model, supersymmetry,
                projective superspace}

\author{Masato Arai}{
  address={High Energy Physics Division, Department of Physical Sciences\\
University of Helsinki and Helsinki Institute of Physics\\
P.O. Box 64, FIN-00014, Finland}
}

\author{Sergei M. Kuzenko}{
  address={School of Physics M013, The University of Western Australia\\
35 Stirling Highway, Crawley W.A. 6009, Australia}
}

\author{Ulf Lindstr\"om}{
  address={Department of Theoretical Physics, Uppsala University\\ 
  Box 803, SE-751 08 Uppsala, Sweden}
  ,altaddress={Helsinki Institute of Physics,\\
 P.O. Box 64, FIN-00014, Finland} 
}

\begin{abstract}
We review the 
projective-superspace
construction of four-dimensional ${\cal N}=2$ 
supersymmetric  sigma 
models on (co)tangent bundles of the  classical 
Hermitian symmetric spaces.
\end{abstract}

\maketitle

Supersymmetry in sigma models is closely related to 
 the geometry of target space \cite{Zumino}. 
In $\cN=1$ models in four space-time dimensions, the target space
 manifold must be K\"ahler \cite{Zumino}.
General $\cN=1$ sigma models can be written in terms of a 
 K\"ahler potential  depending on
 $\cN=1$ chiral superfields and their conjugates.
4D $\cN =2$ models  require the target space geometry to be hyperk\"ahler
 \cite{Alvarez-Gaume:1981hm}. 
Hyperk\"ahler metrics are difficult to construct explicitly.
Thus, manifest $\cN =2$ formulations are needed in order to generate 
$\cN=2$  sigma models 
and, therefore,  hyperk\"ahler metrics.
Projective superspace \cite{Karlhede:1984vr, Hitchin:1986ea}
 provides such a formulation, and it is thus suitable for the construction.
 
 Projective superspace formulation has led to the discovery of 
 several  multiplets that can be used to construct new hyperk\"ahler 
 metrics. One of the most interesting projective multiplets is the so-called
polar multiplet \cite{LR1,LR2} (see also \cite{G-RRWLvU}), for
it can be used to describe a charged 
 $U(1)$ hypermultiplet coupled to a vector multiplet \cite{LR2},
 and therefore is analogous to the $\cN=1$ chiral superfield.
The polar multiplet is described by an arctic superfield $\U(\z)$ and 
 its complex conjugate
 composed with the antipodal map $\bar{\zeta}\rightarrow -1/\zeta$, 
 the antarctic superfield $\breve{\U}(\z)$. 
It is required to possess certain holomorphy properties on a punctured 
 two-plane parametrized by the
 complex variable $\z$ (the latter may be interpreted as a projective
 coordinate on ${\bf C}P^1$).
When realized in ordinary ${\cal N}=1$ superspace, 
 $\U(\z)$ and $\breve{\U}(\z)$ are generated by an infinite
 set of ordinary superfields:
\be
 \U (\z) = \sum_{n=0}^{\infty}  \, \U_n \z^n = 
\F + \S \,\z+ O(\z^2) ~,\qquad
\breve{\U} (\z) = \sum_{n=0}^{\infty}  \, {\bar
\U}_n
 (-\z)^{-n}~.
\label{exp}
\ee
Here $\F$ is chiral, $\S$ complex linear, 
\be
{\bar D}_{\dot{\a}} \F =0~, \qquad \qquad {\bar D}^2 \S = 0 ~,
\label{chiral+linear}
\ee
 and the remaining component superfields are unconstrained complex 
 superfields.  Sigma model couplings of several polar multiplets
 are described by an  action of the form \cite{LR1}
\bea
S[\U, \breve{\U}]  =  
\frac{1}{2\pi {\rm i}} \, \oint \frac{{\rm d}\z}{\z} \,  
 \int {\rm d}^8 z \, 
\cK \big( \U(\z), \breve{\U} (\z), \z  \big) ~.
\label{s-m} 
\eea

It was observed in \cite{K,GK1} that there exists a large subclass in the family of models 
(\ref{s-m}) with interesting geometric properties. 
It corresponds to the case when the Lagrangian $\cK$ in (\ref{s-m}) has no explicit 
$\z$-dependence, 
\bea
S[\U, \breve{\U}]  =  
\frac{1}{2\pi {\rm i}} \, \oint \frac{{\rm d}\z}{\z} \,  
 \int {\rm d}^8 z \, 
K \big( \U^I (\z), \breve{\U}^{\bar{I}} (\z)  \big) ~,
\label{nact} 
\eea
and then $K$ can be interpreted as the  K\"ahler potential of a K\"ahler manifold $\cM$.
Such a  theory
 occurs as a minimal $\cN=2$ extension of the
 general four-dimensional $\cN=1$ supersymmetric nonlinear sigma model \cite{Zumino}
\be
S[\F, \bar \F] =  \int {\rm d}^8 z \, K(\Phi^{I},
 {\bar \Phi}{}^{\bar{J}})  ~.
\label{nact4}
\ee
The target space of the $\cN=2$ sigma model (\ref{nact}) turns out to be 
(an open domain of the zero section) of the cotangent bundle of $\cM$ \cite{GK1}.
By construction,  the model (\ref{nact}) involves an infinite set of auxiliary superfields. 
The hard technical problem is to eliminate these auxiliaries, and then dualize the complex 
linear superfields $\S$'s into chiral ones, 
in accordance with the generalized Legendre transform \cite{LR1}.
This problem was solved in \cite{GK1,AN} for the case 
$\cM= {\bf C}P^n$, the complex projective space, 
and incomplete results were also obtained in \cite{AN} for the complex quadric 
$SO(n+2)/SO(n)\times SO(2)$. 
Recently we have solved the problem for a large class of classical
Hermitian symmetric spaces \cite{AKL}.
In what follows,  a brief review of our construction is presented.

The extended supersymmetric  sigma model  (\ref{nact}) 
inherits  all the geometric features of
its $\cN=1$ predecessor (\ref{nact4}). 
The K\"ahler invariance of the latter,
$K(\F, \bar \F)  \to K(\F, \bar \F)+ \L(\F)+  {\bar \L} (\bar \F) $
turns into 
$K(\U, \breve{\U})  \to K(\U, \breve{\U}) +
\L(\U) + {\bar \L} (\breve{\U} ) $
for the model (\ref{nact}). 
A holomorphic reparametrization of the K\"ahler manifold,
$ \F^I   \to f^I \big( \F \big) $,
has the following
counterpart
$\U^I (\z)  \to
f^I \big (\U(\z) \big)$
in the $\cN=2$ case. Therefore, the physical
superfields of the 
${\cal N}=2$ theory
\be
 \U^I (\z)\Big|_{\z=0} ~=~ \F^I ~,\qquad  \quad \frac{ {\rm d} \U^I (\z) 
}{ {\rm d} \z} \Big|_{\z=0} ~=~ \S^I ~,
\label{kahl4} 
\ee
should be regarded, respectively, as  coordinates of a point in the K\" ahler
manifold and a tangent vector at  the same point. 
Thus the variables $(\F^I, \S^J)$ parametrize the tangent 
 bundle $T\cM$ of the K\"ahler manifold $\cM$ \cite{K}.

The presence of auxiliary superfields $\U_2,~\U_3\,\dots$
 in (\ref{exp}) makes ${\cal N} = 2$ supersymmetry manifest, 
 but the physical content of the theory is hidden.
To describe the theory in terms of 
 the physical superfields $\F$ and $\S$ only, all the auxiliary 
 superfields have to be eliminated  with the aid of the 
 corresponding algebraic equations of motion
\bea
\oint \frac{{\rm d} \z}{\z} \,\z^n \, \frac{\pa K(\U, \breve{\U} 
) }{\pa \U^I} ~ = ~ \oint \frac{{\rm d} \z}{\z} \,\z^{-n} \, \frac{\pa 
K(\U, \breve{\U} ) } {\pa \breve{\U}^{\bar I} } ~ = ~
0 ~, \qquad n \geq 2\,.               
\label{asfem}
\eea
 In general, these equations can be solved only pertubatively.
However, as outlined in \cite{GK1} and elaborated in detail in \cite{AKL}, 
the auxiliary fields may be eliminated exactly
 for any Hermitian symmetric space $\cM$.
 Let us sketch the procedure of constructing such a solution 
 $\U_*(\z) \equiv \U_*( \z; \F, {\bar \F}, \S, \bar \S )$
under the initial conditions (\ref{kahl4}).

Given an arbitrary  point $p_0 \in \cM$, 
there exists a K\"ahler normal coordinate frame 
with origin at $p_0$ such that 
\bea
K_{I_1\dots I_m  \,{\bar J}_1 \dots {\bar J}_n}\Big|_{\F=0} =0~,
\qquad m \neq n~
\eea
and therefore $K(\F, \bar \F) = F (\F \,\bar \F),$
with $F (\F \,\bar \F)$ a real analytic function.
Now, for any  tangent vector $\S_0$ at the origin, 
one can see that  
\bea
\U_0  (\z) = \zeta\S_0
, \qquad  
\breve{\U}_0 (\z) =-  \frac{\bar{\S}_0}{\z}
\label{geo3} 
\eea
solve the equations (\ref{asfem}) at $\F=0$. 
We set $ \U_*( \z; \F =0, {\bar \F} =0 , \S_0, {\bar \S}_0 ) = \zeta\S_0$.
A  next step is to distribute this solution to any point $\F$
of the manifold $\cM$, that is to  make use 
of $ \U_*( \z; \F =0, {\bar \F} =0 , \S_0, {\bar \S}_0 )$ 
in order to obtain $\U_*( \z; \F, {\bar \F}, \S, \bar \S )$.

Let $G$ be the isometry group of $\cM$.
It acts transitively on $\cM$ by holomorphic transformations. 
Let $U$ be the open domain  
on which the normal coordinate system is defined. 
It can  always be chosen such that 
we can construct 
a coset representative, $\cS$:  $U \to  G$, 
where $\cS(p)$:  $\cM \to \cM $
is a holomorphic isometry transformation 
with the property 
$$
\cS(p) \, p_0 =p~, \qquad \cS(p) \in G~,\qquad \forall p \in U~.
$$ 
In other words, $\cS(p) $ maps the origin to $p$.
In local coordinates, $\cS(p) = \cS(\F, {\bar \F})$, and it acts on a generic point 
 $q \in U$ parametrized by complex variables $(\J^I , {\bar \J}^{\bar J} )$ as follows:
\bea
\J \to \J' = f (\J; \F, \bar \F )~, \qquad 
f (0; \F, \bar \F ) =\F~.
\eea
It is crucial 
that the holomorphic isometry transformations 
leave the equations  (\ref{asfem}) invariant.  
This means that applying 
$\cS(\F, \bar \F )$ 
to $\U_0(\z)$, eq. (\ref{geo3}), gives 
\bea 
\U_0(\z) ~\to ~ \U_*(\z) =  f (\U_0(\z) ; \F, \bar \F )
= f (\S_0 \,\z ; \F, \bar \F )~, \qquad 
 \U_*(0)=\F~. \label{sol-trans}
 \eea
Imposing the second initial condition in (\ref{kahl4}),
\be
\S^I =\S^J_0 \, \frac{\pa 
}{\pa \J^J}
f^I (\J; \F, \bar \F ) \Big|_{\J=0} ~,
\label{generalizedlinear}
\ee
we are in a position to 
uniquely express $\S_0$ in terms of $\S$ and $\F$, $\bar \F$.
By construction, $\S$ is a complex linear superfield constrained   
as in (\ref{chiral+linear}). As to $\S_0$, it obeys a generalized 
linear constraint that follows from (\ref{generalizedlinear})
by requiring ${\bar D}^2 \S=0$.

Once the solution $\U_*(\z) \equiv \U_*( \z; \F, {\bar \F}, \S, \bar \S )$ is given, 
we are in a position in principle to compute
the tangent bundle action
\bea
 S_{{\rm tb}}[\F, \bar \F, \S, \bar \S] 
=  
\frac{1}{2\pi {\rm i}} \, \oint \frac{{\rm d}\z}{\z} \,  
 \int {\rm d}^8 z \, 
K \big( \U_* (\z), \breve{\U}_* (\z)  \big) ~.
\label{tb-act}
\eea
But it is extremely important to choose a simplest coset representative
(using the natural freedom in its choice
$\cS(\F, {\bar \F}) \to \cS(\F, {\bar \F})\,h(\F, {\bar \F})$, 
with $h(\F, {\bar \F})$  taking its values in the isotropy subgroup  $H$ at $p_0$).
With a complicated coset representative chosen, 
it will be practically impossible to do the contour integral on the right of 
(\ref{tb-act}).
Finally, after having evaluated  the contour integral in
(\ref{tb-act}), it only remains to dualize the complex linear tangent variables
$\S$'s into chiral one-forms, in accordance with the generalized Legendre 
transform \cite{LR1}. The target space for the model obtained is 
$T^*\cM$ in the compact case, and a part of $T^*\cM$ in the non-compact case.
In our work \cite{AKL}, 
the above procedure was carried out for the following Hermitian symmetric spaces
(HSS):
\bea
\left(\begin{array}{c|c|c|c|c}
 \mbox {compact HSS} & \frac{U(n+m)}{U(n)\times U(m)} & \frac{SO(2n)}{U(n)}&
 \frac{Sp(n)}{U(n)} &\frac{SO(n+2)}{SO(n)\times SO(2)}\\
  \hline 
  \mbox{non-compact HSS}& \frac{U(n,m)}{U(n)\times U(m)}  &
   \frac{SO^*(2n)}{U(n)} &  \frac{Sp(n, {\bf R}) }{U(n)}  &\frac{SO_0(n,2)}{SO(n)\times SO(2)}
  \end{array}
\right) \non 
\eea

As an example, consider the 
 Grassmannian $\cM= G_{m,n+m}$. Its K\"ahler potential is
\be
K (\F , \F^\dagger )= \ln \det ( {\bf 1}_m + \F^\dagger \F) 
= \ln \det ({\bf 1}_n + \F \F^\dagger )~.
\label{Kalpot}
\ee
where $\F= (\F^{i \a})(i=1,\dots,n,~\alpha=1,\dots,m)$. 
The useful coset representative is
\bea
\cS(\F, \bar \F) = \left(
\begin{array}{cc}
\underline{s} ~  ~& ~\F\,s\\
-\F^\dagger \, \underline{s}  ~~& ~s 
\end{array}
\right)~, \qquad s^{-2}=\F^\dagger \F + {\bf 1} _m ~, 
\quad \underline{s}^{-2} = \F \F^\dagger  + {\bf 1} _n~.
\label{cosetrep3}
\eea
 Its use leads to the tangent bundle action \cite{AKL}
\bea
 S_{\rm tb}= \int {\rm d}^8z \left\{
 K(\F , \F^\dagger) 
+ \ln \det \Big( {\bf 1} _m
-({\bf 1} _m +\Phi^\dagger \Phi)^{-1} \Sigma^\dagger ({\bf 1} _n + \F \Phi^\dagger )^{-1}
\Sigma \Big) \right\} ~.
\label{action-gr}
\eea
Finally, applying the generalized Legendre 
transform   to (\ref{action-gr}) gives the $\cN=2$ sigma model 
originally constructed in \cite{Lindstrom:1983rt}.


\bibliographystyle{aipproc}   

\end{document}